\documentclass[usegraphicx,usenatbib]{mn2e}

\usepackage[total={17.8cm,24.0cm},centering]{geometry}
\usepackage{times}
\usepackage{amssymb}

\newcommand{\apj}{ApJ}           
\newcommand{\apjl}{ApJ}           
\newcommand{\mnras}{MNRAS}       
\newcommand{\nat}{Nature}
\newcommand{\aap}{A\&A}
\newcommand{\araa}{ARA\&A}
\newcommand{\aj}{AJ}
\newcommand{\pasp}{PASP}
\newcommand{\apjs}{ApJS}           
\newcommand{\aapr}{A\&A Rev.}

\newcommand{\kms}{\hbox{km s$^{-1}$}}
\newcommand{\msun}{\hbox{$M_\odot$}}
\newcommand{\lsun}{\hbox{$L_\odot$}}
\newcommand{\ml}{\hbox{$M/L$}}
\newcommand{\re}{\hbox{$R_{\rm e}$}}
\newcommand{\bh}{\hbox{$M_{\rm BH}$}}

\newcommand{\vs}{\hbox{$V/\sigma$}}
\newcommand{\vse}{\hbox{$(V/\sigma,\varepsilon)$}}
\newcommand{\plotone}[1]{\includegraphics[width=\columnwidth]{#1}}
\newcommand{\refsec}[1]{Section~\ref{#1}}
\newcommand{\reffig}[1]{Fig.~\ref{#1}}

\title[The black hole in Centaurus~A]
{The mass of the black hole in Centaurus~A from SINFONI AO-assisted integral-field observations of stellar kinematics}

\author[Cappellari et al.]
{Michele Cappellari,$^1$\thanks{E-mail: cappellari@astro.ox.ac.uk}
N. Neumayer,$^2$
J. Reunanen,$^3$
P.~P. van der Werf,$^4$\newauthor
P.~T. de Zeeuw$^{2,4}$
and H.-W. Rix$^5$\\
$^1$ Sub-Department of Astrophysics, University of Oxford, Denys Wilkinson Building, Keble Road, Oxford OX1 3RH\\
$^2$European Southern Observatory, Karl-Schwarzschild-Str~2, 85748 Garching, Germany\\
$^3$Tuorla Observatory, University of Turku, V\"ais\"al\"antie 20, 21500 Piikki\"o, Finland\\
$^4$Leiden Observatory, Leiden University, P.O.\ Box 9513, 2300 RA Leiden, The Netherlands\\
$^5$Max-Planck Institute for Astronomy, K\"onigstuhl 17, 69117
  Heidelberg, Germany}

\date{Accepted for publication in MNRAS}

\pagerange{\pageref{firstpage}--\pageref{lastpage}} \pubyear{2008}

\begin{document}
\label{firstpage}
\maketitle

\begin{abstract}
We present a determination of the mass of the supermassive black hole (BH) and the nuclear stellar orbital distribution of the elliptical galaxy Centaurus~A (NGC~5128) using high-resolution integral-field observations of the stellar kinematics. The observations were obtained with SINFONI at the ESO Very Large Telescope in the near-infrared ($K$-band), using adaptive optics to correct for the blurring effect of the earth atmosphere. The data have a spatial resolution of 0\farcs17 FWHM and high $S/N\ga80$ per spectral pixel so that the shape of the stellar line-of-sight velocity-distribution can be reliably extracted. We detect clear low-level stellar rotation, which is counter-rotating with respect to the gas. We fit axisymmetric three-integral dynamical models to the data to determine the best fitting values for the BH mass $\bh=(5.5\pm3.0)\times10^7 \msun$ ($3\sigma$ errors) and $(\ml)_K=(0.65\pm0.15)$ in solar units. These values are in excellent agreement with previous determinations from the gas kinematics, and in particular with our own published values, extracted from the same data. This provides one of the cleanest gas versus stars comparisons of \bh\ determination, due to the use of integral-field data for both dynamical tracers and due to a very well resolved BH sphere of influence $R_{\rm BH}\approx0\farcs70$. We derive an accurate profile of the orbital anisotropy and we carefully test its reliability using spherical Jeans models with radially varying anisotropy. We find an increase in the tangential anisotropy close to the BH, but the spatial extent of this effect seems restricted to the size of $R_{\rm BH}$ instead of that $R_b\approx3\farcs9$ of the core in the surface brightness profile, contrary to detailed predictions of current simulations of the binary BH scouring mechanism. More realistic simulations would be required to draw conclusions from this observation.
\end{abstract}


\section{Introduction}

The existence of supermassive black holes (BHs) in normal galaxy nuclei was predicted forty years ago by \citet{LyndenBell69}, but until fifteen years ago it was still considered an interesting possibility which had to be demonstrated. Nowadays BHs are regarded as a crucial ingredient for our understanding of how galaxies form. Key to this paradigm shift was the launch in 1990 of the Hubble Space Telescope (HST). It all started with the realisation that the mass of the BH is correlated to other global characteristics of the host galaxy as a whole. Initially a correlation  $\bh-L$ was found between the mass of the BH and the luminosity of the host-galaxy stellar spheroid \citep{kormendy95,magorrian98}. In 1997 the installation of the STIS long-slit spectrograph on HST allowed the spatially-resolved kinematical observations to probe inside the radius of the subarcsecond BH sphere of influence $R_{\rm BH}\equiv G \bh/\sigma^2$ in nearby galaxies ($\sigma$ being the velocity dispersion of the stars in the galaxy). The increased accuracy in the \bh\ determinations contributed to the discovery of the tighter $\bh-\sigma$ correlation \citep{ferrarese00,gebhardt00}.

As $\sigma$ is a good predictor of galaxy properties, it is perhaps not surprising that similar correlations were found between \bh\ and respectively the galaxy concentration \citep{graham01}, the dark-halo mass \citep{ferrarese02,Baes03,pizzella05}, the bulge mass \citep{McLure02,marconi03,haring04} and the stars' gravitational binding energy \citep{aller07}. The existence of these correlations is broadly consistent with a scenario in which the BH regulates the galaxy formation, during the hierarchical galaxy merging, by shutting off the conversion of gas into stars via a feedback mechanism due to its powerful outflows \citep{silk98,granato04,dimatteo05,Bower06}.

Even though the current scenario explains a number of observed facts, our interpretation of the role of BHs in galaxy formation is far from secure. One of the problems lies in the fact that the models have few observables to compare with, mainly the galaxy mass (or $\sigma$) and \bh. Moreover, even after a decade of HST spectroscopy and models, only about 40 secure BH determinations exist \citep[compilations are given e.g. in][]{tremaine02,ferrarese05rev,Graham08}. With few exceptions the BH determinations have been performed with a single dynamical tracer (either gas or stars), so that no independent test of the two measurements methods could be made. Very little is known about the orbital distribution near the BHs, which is expected to contain key information on the BH accretion process \citep{quinlan97,milosavljevic01}.

The advent of integral-field spectroscopy on all the 8--10-m class telescopes, combined with adaptive optics (AO) to reduce the blurring effect of the Earth's atmosphere is opening a new epoch for BH studies. The large mirrors allow for shorter exposure times and higher signal-to-noise ratios ($S/N$) of the observations, than what was possible with the 2.4-m mirror and the STIS long-slit spectrograph of HST. Observations at near-infrared wavelengths allow dust absorption effects to be virtually eliminated. The integral-field observations, due to the tight constraint on the orbital distribution, dramatically improve the accuracy of \bh\ determination, for a given spatial resolution and $S/N$ \citep{verolme02}. Integral-field data are also needed for a unique recovery of the orbital distribution from the observations \citep{cappellari05,krajnovic05,vandeVen08}. This can be understood from dimensional arguments, considering that most orbits in a stationary potential conserve three isolating integrals of motion. This three-dimensional orbital distribution cannot be recovered without the knowledge of at least another three-dimensional observed quantity. Motivated by these arguments, \bh\ determinations from AO-assisted integral-field observations in the near-infrared are starting to appear in the literature \citep{Davies06,nowak07,Nowak08}.

The elliptical galaxy NGC~5128 (Centaurus~A) is a prime candidate for AO-assisted integral-field observations. At least nine independent distance $D$ determinations for Cen~A, based on different methods, are available in the literature\footnote{See the NED-1D compilation by Barry F. Madore and Ian P. Steer at http://nedwww.ipac.caltech.edu/level5/NED1D/.} \citep[e.g.][]{tonry01,rejkuba04,ferrarese07}. The median value is $D=3.8$ Mpc, with extreme ranges of 3.4 Mpc and 4.4 Mpc respectively. Here we adopt a value\footnote{The choice of the distance $D$ does not affect our results but sets the scale of our models in physical units. Specifically, lengths and masses scale as $D$, while mass-to-light ratios scale as $D^{-1}$.} $D=3.5$ Mpc to be consistent with all the earlier papers on \bh\ determination on this galaxy \citep{marconi01,marconi06,silge05,haring06,krajnovic07,Neumayer07}. At this close distance Cen~A is the nearest elliptical galaxy and one arcsec corresponds to 17 pc.

Cen~A is among the only $\sim10$  galaxies on the whole sky with an observed $R_{\rm BH}\approx1\arcsec$ \citep[see][for a partial list]{kormendy04}. Moreover Cen~A possesses a nuclear gaseous disk in regular rotation from which a number of independent determinations of \bh\ have been performed, with ever increasing accuracy \citep{marconi01,marconi06,haring06,krajnovic07,Neumayer07}. This allows for an accurate comparison between the BH mass derived with gaseous or stellar kinematics. The $K$-band central surface brightness of Cen~A is quite bright at $\mu_K\approx12.2$ mag \citep{Jarrett03}, which allows a high $S/N$ in the stellar spectra to be achieved in reasonable exposure times. A bright star, sufficiently close to the nucleus, can be used as reference for the AO correction. All these facts make Cen~A a unique observational benchmark for \bh\ determinations in the near-infrared.

In Section~2 we describe our data and the extraction of the stellar kinematics. In Section~3 we present the stellar dynamical models. We discuss our results in Section~4 and we finally summarize them in Section~5.

\section{Observations and data analysis}

\subsection{Spectroscopic data}

For the dynamical modeling of the stellar kinematics we used integral-field spectroscopy obtained with SINFONI on the UT4 (Yepun) of the Very Large Telescope of the European Southern Observatory on the Cerro Paranal.  SINFONI consists of the cryogenic near-IR integral field spectrometer SPIFFI \citep{Eisenhauer03a,Eisenhauer03b} coupled to the visible curvature AO system MACAO \citep{Bonnet03}. We observed the nucleus of Cen~A with two different spatial scales: 0\farcs250$\times$0\farcs125 (250mas scale) with a Field of View (FoV) of 8\arcsec$\times$8\arcsec and 0\farcs10$\times$0\farcs05 (100mas scale) with a FoV of 3\farcs2$\times$3\farcs2. The 100mas observations and data reduction were already described in \citet{Neumayer07}. The observations were performed in excellent seeing conditions of 0\farcs5 FWHM, as measured by the seeing monitor in the $V$-band. With the 250mas spatial scale the observations were taken in natural seeing, while with the 100mas scale the SINFONI AO module was locked onto an R$\sim$14 mag star about 36\arcsec southwest of the nucleus. Although the reference star is relatively distant from the galaxy nucleus, the good seeing allowed us to achieve a nearly diffraction-limited correction in the $K$-band. Both data sets were obtained in the $K$-band, which covers the wavelength range 1.93--2.47\micron. The spectral resolution of the observations was $R\sim4800$ in both scales, and corresponds to an instrumental dispersion of $\sigma_{\rm ins}\sim27$ \kms.

The 250mas observations followed an Object-Sky-Object sequence with equal exposure time of 300 s each, for a total on-source exposure time of $4\times300=1,200$ s. The sky exposure were taken at 200\arcsec from the nucleus to make sure the spectra were not contaminated by the large galaxy. The 100mas observations followed a similar sequence with individual exposure times of 900 s each and a total on-source exposure time of $15\times900=13,500$ s. The different exposures were dithered with shifts of 0\farcs2 to allow for the removal of detector defects and cosmic rays. The data were reduced using the SINFONI data reduction pipeline provided by ESO as described in section~2 of \citet{Neumayer07}.

The spatial point-spread-function (PSF) of the 100mas observation was determined in section~2.1 of \citet{Neumayer07} from a fit to the non-thermal nucleus, which is unresolved down to a 0\farcs06 (FWHM) spatial resolution \citep{haring06,marconi06}. The PSF can be approximated by two Gaussians with FWHM of 0\farcs12 and 0\farcs30 respectively, with the smallest Gaussian containing 17\% of the total flux. A two-dimensional fit with a single Gaussian to the nucleus in the reconstructed image from the data cube gives a FWHM of 0\farcs17.

For the stellar dynamical modeling the integral-field high-spatial resolution SINFONI observations are essential to tightly constrain the BH mass and stellar orbital distribution. However they are not sufficient as they sample only a small fraction of the  half-light radius of Cen~A  ($\re\approx83\arcsec$ in $K$-band; \citealt{Jarrett03}). \citet{shapiro06} showed that one has to sample with the kinematics a significant fraction of \re\ to accurately constrain BH masses. For this reason in this paper we also use the $K$-band kinematics obtained with the Gemini Near Infrared Spectrograph (GNIRS) at Gemini South by \citet{silge05}. The long-slit observations extend to a distance of $R\approx40\arcsec$ from the galaxy nucleus and were obtained at two position angles: (i) centered on the nucleus, along the galaxy major axis at large radii (${\rm PA_{phot}}\approx35^\circ$ from North through East), and (ii) along the galaxy minor axis (${\rm PA_{phot}}\approx125^\circ$), but offset from the nucleus by 0\farcs85. In this paper we increased the GNIRS $\sigma$ by 5\% to match our SINFONI kinematics. Differences  at this level between the two datasets are almost unavoidable and are likely due to low-level systematic calibration errors, or differences in the adopted stellar templates. For this reason we cannot state which of the two datasets has the proper absolute $\sigma$ calibration. As none of the results in this paper depends on whether we apply the small shift to the GNIRS or SINFONI data, we arbitrarily decided to increase the GNIRS data to match our SINFONI ones, which we adopt as reference. A 5\% uncertainty in $\sigma$ approximately translates into a 10\% uncertainty in the $M/L$.

\subsection{Merging and binning}

The individual SINFONI exposures were registered using reconstructed images and merged into a single data cube. In the process they were resampled to a spatial scale of 0\farcs125$\times$0\farcs125 (250mas scale) and 0\farcs05$\times$0\farcs05 (100mas scale).

The stellar kinematics requires a high $S/N$ for an unbiased extraction from the spectra. For typical observational setups, values on the order of $S/N\ga50$ per spectral pixel, are generally required to detect deviations of the line-of-slight velocity distribution from a simple Gaussian shape \citep[e.g.][]{vanDerMarel93,Bender94,Statler95}. For this reason long-slit data are invariably adaptively binned along the spatial direction, before the kinematic extraction \citep[e.g.][]{vanDerMarel94}. The goal of binning is to maintain the maximum spatial resolution, given a constraint on the minimum $S/N$. Four our integral-field data we use the Voronoi binning technique,\footnote{Available from http://www-astro.physics.ox.ac.uk/$\sim$mxc/idl/} which constitutes an optimal solution to this problem in two-dimension \citep{cappellari03}.

Given that the observations are dominated by the photon noise, we spatially binned the data by requiring an equal number of counts in each spatial bin. In this way we did not have to rely on the, often inaccurate, noise propagation by the reduction pipeline, to estimate the $S/N$. This constant flux per bin is achieved in the Voronoi binning algorithm by setting the input noise $N=\sqrt{S}$, where $S$ is the total flux (in arbitrary units) in each unbinned spectrum.  For the 100mas scale the target $S/N$ for the binning was fixed by the requirement for the data to be unbinned (pixel size 0\farcs05$\times$0\farcs05) within a radius $R\la0\farcs2$ and start being binned at larger radii, where the surface brightness decreases. This led to 622 Voronoi bins, out of the original $64^2=4096$ pixels. Similarly for the 250mas scale we required the data to be unbinned (pixel size 0\farcs125$\times$0\farcs125) within $R\la0\farcs5$, leading to 554 bins. The actual minimum $S/N$ achieved by the adopted binning was determined afterwards, from the rms residuals of a spectral template fit (\refsec{sec:ppxf}), to be $(S/N)_{\rm min}\sim80$ for both spatial scales.

\subsection{Extraction of the stellar kinematics}
\label{sec:ppxf}

The near-infrared $K$-band spectral region is dominated by the strong stellar absorption feature of the 2.30\micron\ (2-0) $^{12}$CO band head. At this wavelength the galaxy spectrum is dominated by the light from cool and evolved giant stars. The strength of the CO absorption is very sensitive to the star surface gravity \citep{Kleinmann86,Wallace97} and this implies that, contrary to the optical region, in the $K$-band it is {\em essential} to include giant stars in the construction of an optimal stellar template for an unbiased extraction of the stellar kinematics \citep{silge03}.

The extraction of the stellar kinematics was performed using the penalized pixel-fitting method$^3$ \citep[pPXF;][]{cappellari04}, which fits the logarithmically-rebinned spectra with a template convolved with a line-of-sight velocity-distribution described by a Gauss-Hermite expansion \citep{vanDerMarel93,gerhard93}. The method allows the template to be carefully optimized during the kinematics fit, it permits emission lines to be easily masked, and it includes a penalty criterion to deal with low $S/N$ or insufficient resolution in the spectra.  As library of stellar templates we used a set of 11 dwarfs and giants stars (luminosity class II--V) of late spectral types (K--M), observed with the same instrumental setup as for the Cen~A observations.

\begin{figure}
\centering
  \plotone{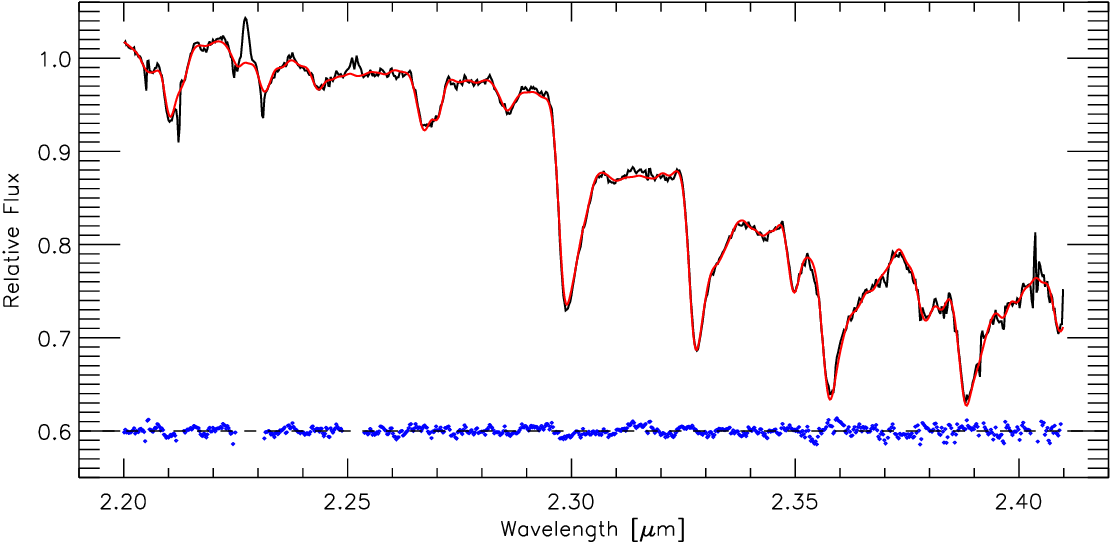}
  \caption{Optimal template for Cen~A. The total galaxy spectrum observed with SINFONI within the 8\arcsec$\times$8\arcsec\ 250mas field (black solid line) is compared to the broadened optimal template determined with pPXF (red solid line). The fit residuals, vertically shifted by an arbitrary amount, are shown as blue dots. The gaps in the residuals correspond to some non well corrected telluric features and two H$_2$ gas emission lines at 2.23\micron\ and 2.25\micron, which were excluded from the fit.}
  \label{fig:template_fit}
\end{figure}

The optimal positive linear combination of the 11 templates was determined only once, using a spectrum with very high $S/N\approx240$, obtained by co-adding all the spectra in the 250mas SINFONI observations, while excluding the unresolved non-stellar nucleus. As expected the flux in the resulting optimal template fit is dominated by a giant M5III star (52\% of the flux). This star provides by itself an excellent fit to the central spectrum of Cen~A. However the contribution from a dwarf M0V star (30\% of the flux) is required to properly reproduce the depth of the Na feature at 2.21\micron\ \citep[cf.][]{Lyubenova08}. The template provides a very accurate description of the spectrum from 2.2\micron--2.4\micron, which contains all the significant stellar absorption features in the $K$-band (\reffig{fig:template_fit}).

For maximum consistency the same combination of stellar templates was used to extract the kinematics at all spatial positions and for both the 250mas and 100mas observations. Additive polynomials of degree four were used to allow for possible variations in the stellar line-strength at different radii, to account for imperfections in the sky subtraction or spectral calibration, and to model the contribution of the non-thermal nucleus (\refsec{sec:nonthermal}). Consistent results were obtained using polynomials of degree 1--8. From numerous tests on other galaxies we found that, even for the most extreme gradients in the stellar population, the additive polynomials are sufficient to approximate possible low-frequency variations in the line-strength, once the global galaxy template is accurately determined from a high-$S/N$ spectrum, to account for the high-frequency content in the spectrum. We prefer this faster and more robust approach than trying to fit the template mix in every bin, from spectra of lower $S/N$. Generally both approaches give nearly indistinguishable results, and we tested that this is true also in the present case.

\begin{figure}
\centering
  \plotone{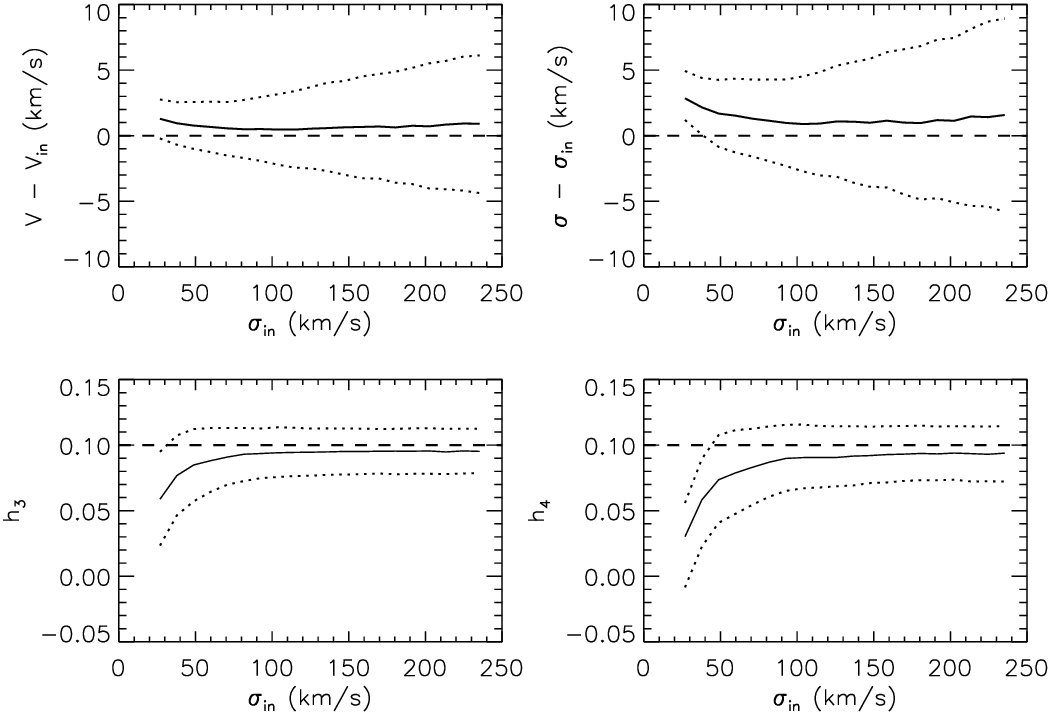}
  \caption{Testing of penalization in pPXF. We simulated spectra with the $S/N=80$ of our data and an LOSVD with $h_3=0.1$, $h_4=0.1$ and $\sigma$ in the range between 30 and 240 \kms. We extracted the kinematics with pPXF and a penalty $\lambda=0.6$. The lines in the top two panels show the 50th (median, solid line), 16th and 84th percentiles (1$\sigma$ errors, dotted lines) of the differences between the measured values and the input values of the mean velocity $V_{\rm in}$ and the velocity dispersion $\sigma_{\rm in}$. The bottom panels show the same lines for the recovered values of $h_3$ and $h_4$, compared to the input values (dashed line). In the range of our observations $\sigma\ga130$ \kms\ the recovered values are virtually unbiased.}
  \label{fig:ppxf_simulation}
\end{figure}

We fitted with pPXF the mean velocity $V$, the velocity dispersion $\sigma$ and the first two Gauss-Hermite parameters $h_3$ and $h_4$ \citep{vanDerMarel93,gerhard93}, using the same spectral range and masked regions as in the real data. We adopted a penalty $\lambda=0.6$ (defined in equation~12 of \citealt{cappellari04}). The LOSVD is always very well sampled by our observations and the $S/N$ is high, so any $\lambda\la0.6$ gives essentially the same results and the measurements are virtually unbiased (\reffig{fig:ppxf_simulation}). The $1\sigma$ measurement errors were determined as the biweight dispersion \citep{Hoaglin83} of 100 Monte Carlo realizations. For the determination of realistic errors we used non-penalized fits with $\lambda=0$. At our minimal $S/N\approx80$ and for $\sigma\approx150$ \kms\ the typical random errors in $V$, $\sigma$, $h_3$ and $h_4$, are 4 \kms, 5 \kms, 0.02 and 0.03 respectively. The kinematics extraction was performed while fitting precisely the same wavelength region, from 2.25\micron--2.37\micron, in both the 250mas and 100mas SINFONI observations (\reffig{fig:spectrum_variation}). This spectral region includes the (2--0) $^{12}$CO band head and three other prominent band heads of the $^{12}$CO series. Emission from the highly ionized species of [CaVIII] at 2.32\micron\ and an H$_2$ line at 2.35\micron\ were excluded from the fits \citep[see also][]{Riffel08}.

\begin{figure}
  \plotone{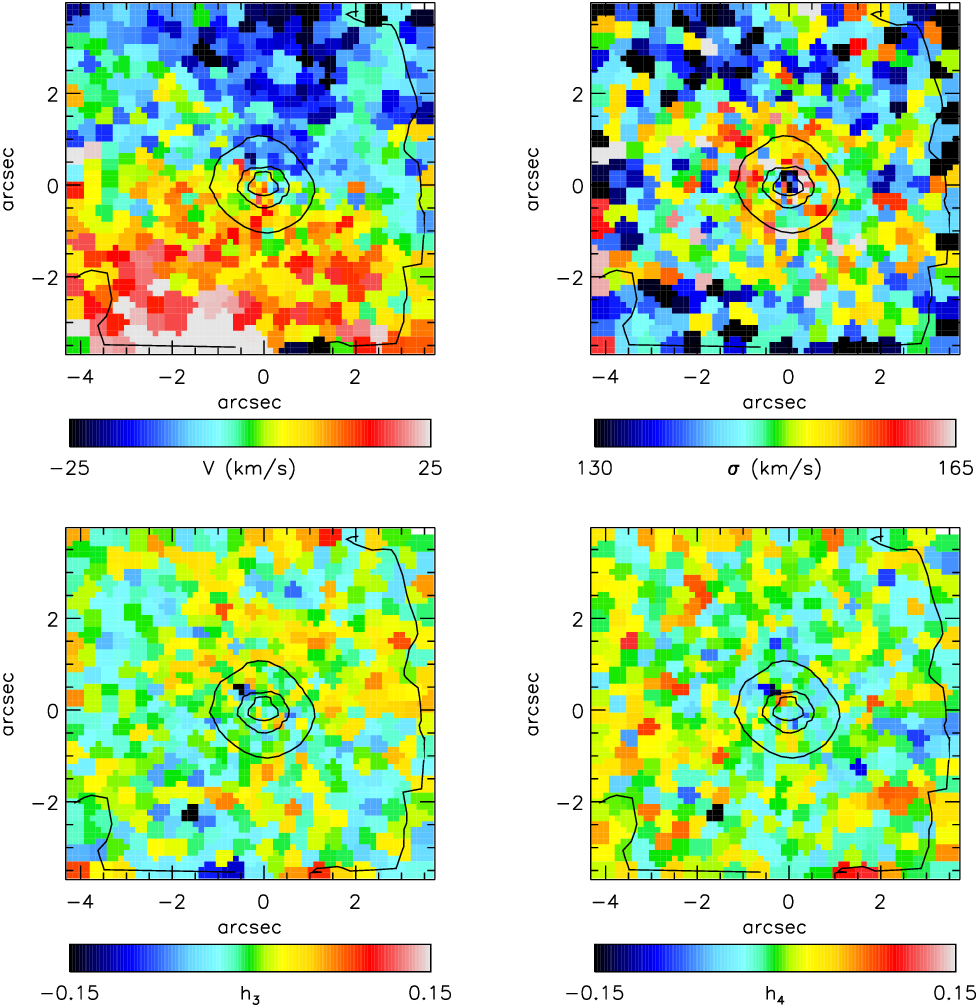}
  \caption{Voronoi binned stellar kinematics from the SINFONI 250mas observations. The four panels show the mean velocity $V$, the velocity dispersion $\sigma$ and the two Gauss-Hermite moments $h_3$ and $h_4$. Overlaid are the contours of the surface brightness derived from the reconstructed image, in 1 mag intervals. North is up and East is to the left.}
  \label{fig:kinematics250}
\end{figure}

\begin{figure}
  \plotone{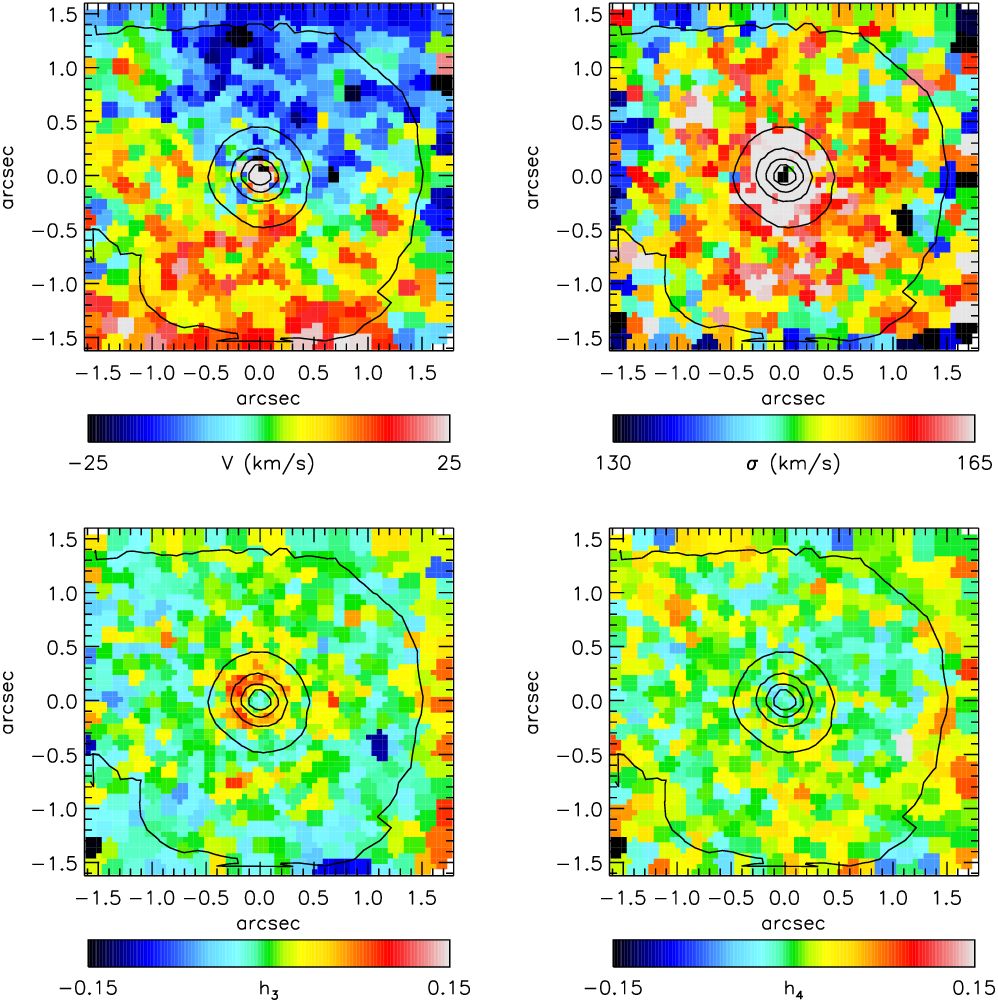}
  \caption{Same as in \reffig{fig:kinematics250} for the SINFONI 100mas observations. The color levels are the same in this figure and in \reffig{fig:kinematics250}.}
  \label{fig:kinematics100}
\end{figure}

The extracted SINFONI kinematics\footnote{Available from http://www-astro.physics.ox.ac.uk/$\sim$mxc/cena09.} in the two scales is shown in \reffig{fig:kinematics250} and \reffig{fig:kinematics100}. For the first time our observations detect a low-level clear sense of stellar rotation in Cen~A.  We used the method of Appendix~C of \citet{krajnovic06} to determine the global kinematical major axis ${\rm PA_{kin}}=167^\circ\pm8^\circ$ and the systemic velocity of $V_{\rm syst}=531\pm5$ \kms. A maximum velocity of just $\Delta V\approx20$ \kms\ is reached along ${\rm PA_{kin}}$ at about $R\approx4\arcsec$ from the nucleus, at the edge of our 250mas SINFONI FoV.
The $\sigma$ field shows a gradual increase towards the center, before a sudden drop in the very center, where the non-thermal continuum prevents a proper kinematics extraction. The $h_3$ field is symmetric about the center, indicating low template-mismatch, and anti-correlated with $V$, as generally observed in early-type galaxies \citep{Bender94,Krajnovic08}, and the $h_4$
field is generally close to zero over the whole field.

The nuclear stellar rotation is {\em counter-rotating} (by about 180$^\circ$) with respect to the regular H$_2$ nuclear gas rotation presented in \citet[their fig.~6]{Neumayer07}. The stars also rotate much slower than the gas, which reaches a maximum velocity $\Delta V\approx130$ \kms\ at $R\approx0\farcs5$. This indicates that the recent gas acquisition was not able to produce a significant fraction of stars near the nucleus. This is consistent with the lack of evidence for any change in the nuclear stellar population of Cen~A. This is different from a similar case of gas versus stars counter-rotation observed in the nucleus of the spiral galaxy NGC~5953 \citep{FalconBarroso06}. In that case the nuclear gas rotation is associated to evidence of young stars which co-rotate with the gas and counter-rotate with respect to the outer part of the galaxy.

\subsection{Contribution from the non-thermal nucleus}
\label{sec:nonthermal}

In the case of Cen~A adopting a fixed template becomes important very close to the center, due to the presence of the non-thermal nucleus \citep{marconi00}. At radii where the nucleus dominates, the stellar absorptions are diluted, resulting in a strong decrease in the observed line-strength $\gamma$. Due to the strong correlation between $\gamma$ and the stellar velocity dispersion $\sigma$ (section~2.2 of \citealt{vanDerMarel93}), an inaccurate modeling of the non-thermal dilution can cause large errors in the measured $\sigma$. Similarly the rise in the non-thermal continuum could be incorrectly interpreted as a variation of the stellar population, requiring a change in the stellar template mix and also producing an error in $\sigma$. As there is no evidence for a sudden change in the population in the nucleus of Cen~A the safest choice is to assume the stellar template is fixed and to model the non-thermal continuum via additive polynomials in pPXF. Additive polynomial still allow for low-order variations in the stellar line-strength and account for possible instrumental effects. This approach allows the $\sigma$ to be reliably extracted in the high-resolution 100mas observations down to $R\ga0\farcs2$, before the photon noise of the nucleus eliminates all stellar information from the spectra.

\begin{figure}
\centering
  \plotone{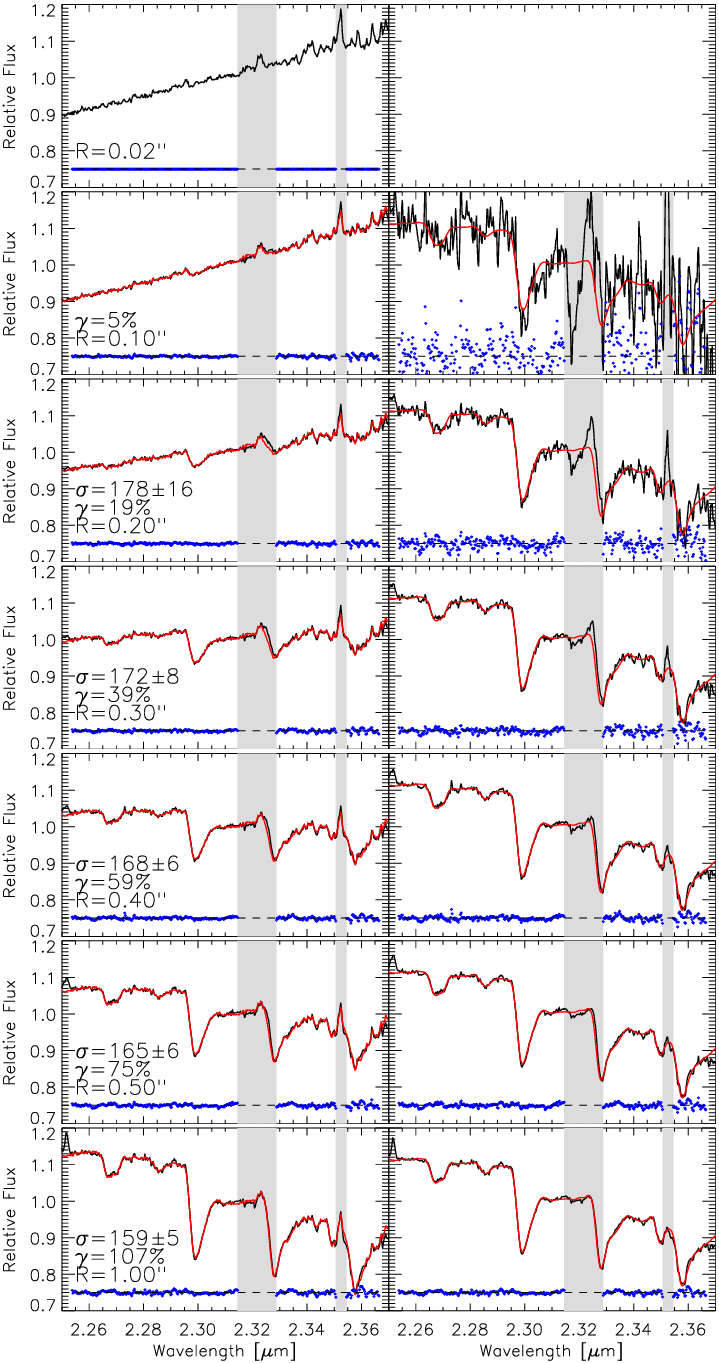}
  \caption{Radial variation in the spectrum of Cen~A in the 100mas SINFONI observations. {\em Left Column:} Different panels show the observed spectra (black solid line) obtained by co-adding the spectra of the spaxels contained within circular annuli of radius $R$ and one pixel width (0\farcs05). The best-fitting pPXF model (red solid line) consists of the stellar template (\reffig{fig:template_fit}) convolved with a Gaussian LOSVD plus a fourth degree additive polynomial, plus a scaled copy of the non-thermal nuclear spectrum (top panel). The residuals are shown at the bottom of each panel with the blue dots. The value $\gamma>100\%$ at $R=1\arcsec$ is likely due to a small difference in the line strength of the stellar spectrum, or in the sky subtraction, with respect to the adopted optimal template.  Emission from the [CaVIII] line at 2.32\micron\ and the H$_2$ line at 2.35\micron\ were excluded from the fits (grey areas). These lines are prominent also outside the nucleus and cannot be properly subtracted by our model. {\em Right Column:} The convolved optimal template (red solid line) is compared to the observed spectrum after subtraction of the nuclear spectrum and polynomial contributions (black solid line). The blue dots show the residuals.}
  \label{fig:spectrum_variation}
\end{figure}

As an illustration of the importance of taking the non-thermal continuum accurately into account in the kinematic extraction, in \reffig{fig:spectrum_variation} we show the best fit with pPXF, using a Gaussian LOSVD, to the spectra extracted at different radii from the 100mas SINFONI data. We also performed separate fits including the Gauss-Hermite parameters $h_3$ and $h_4$. In the range 0\farcs3--1\arcsec, where we can trust our extraction, the LOSVD is essentially consistent with a Gaussian as $h_4=-0.02\pm0.03$, in agreement with \reffig{fig:kinematics100}. The high $S/N$ spectra in these plots were obtained by co-adding all the spectra within circular annuli of one pixel width (0\farcs05). The standard kinematics extraction approach for this paper consists of modeling the observed spectrum as the sum of a convolved fixed optimal template plus a fourth degree additive polynomial. To test the sensitivity of the measured $\sigma$ to the details of the extraction, especially in the critical continuum-dominated nuclear region, we compared our standard approach with others using different options in the pPXF routine: (i) We use a first degree polynomial; (ii) We fit the optimal template at each radius using our 11 stars; (iii) we use as template a single M5III star; (iv) We include in the fit the additive contribution of a scaled version of the nuclear non-thermal spectrum (as described in \citealt{Kelson00}, to subtract the sky spectrum). All these four approaches give dispersion profiles which agree within the statistical errors for $R>0\farcs2$.

The last approach is the one which produces the best fit and describes almost every details of the observations. This is the one we show in \reffig{fig:spectrum_variation}. We also give in the plots the line-strength $\gamma$, defined as the ratio between the mean flux  within the fitted range in the observed Cen~A spectrum, and the flux contributed by the best fitting stellar template alone in the same range. At $R\la0\farcs2$ the non-thermal source dominates and the spectrum shows a nearly linear trend. At $R\approx0\farcs2$ it becomes possible to estimate the velocity dispersion, however the stars still contribute only 19\% of the flux in the observed spectrum, the statistical errors are large and the determination is still affected by large systematic uncertainties. At this radius the measured $\sigma$ is sensitive to the details of the extraction. Only at larger radii the kinematics can be reliably extracted and will be included in our models.

\begin{figure}
  \plotone{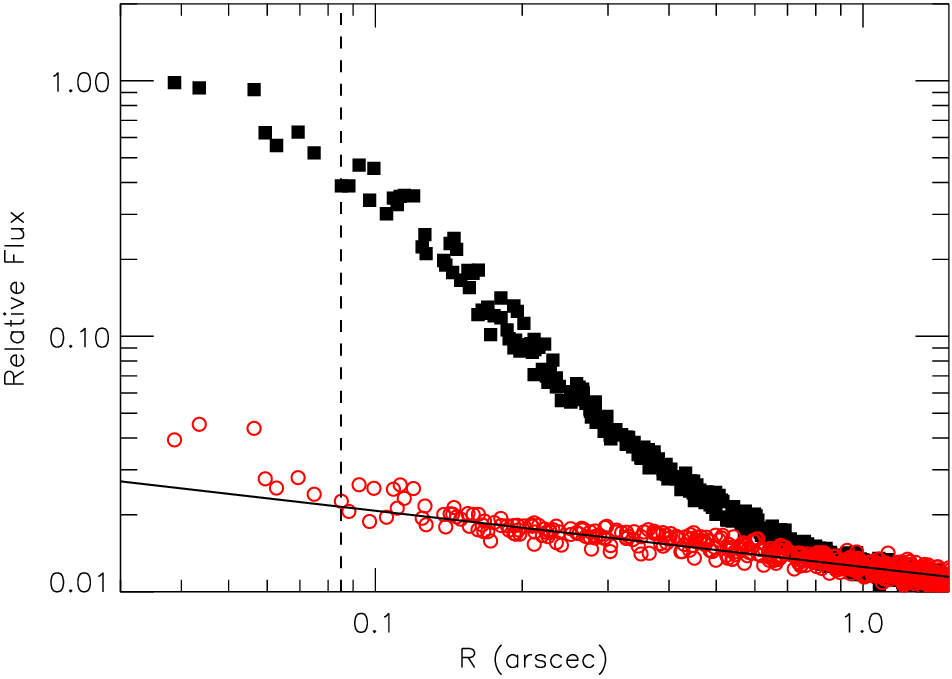}
  \caption{The radial surface brightness profile $I(R)$ as measured from the individual pixels in the SINFONI 100mas scale (filled squares) is compared to the estimated profile of the stellar light $\Sigma(R)=I(R)\gamma(R)$ (red open circles). As expected for a normal galaxy profile, the underlying stellar light profile is smooth and well approximated by a power-law $\Sigma(R)\propto R^{-0.22}$ (solid line). The dashed vertical line marks the radius where the PSF has reached 50\% of its peak value.}
  \label{fig:gamma_profile}
\end{figure}

\reffig{fig:spectrum_variation} shows that the non-thermal continuum is still present in the spectra at the 25\% level at $R\approx0\farcs5$. The effect of the non stellar continuum is still clearly visible in the varying slope of the spectrum. Still at all radii the stellar spectrum is well approximated by the fixed convolved template, once the continuum is removed (right column of \reffig{fig:spectrum_variation}). To better quantify the extent of the dilution due to the central non-thermal continuum \reffig{fig:gamma_profile} shows the surface brightness profile $I(R)$ measured from the reconstructed image of the 100mas data cube, together with the radial profile of the the pure stellar contribution, estimated as $\Sigma(R)=I(R)\gamma(R)$ from the result of the pPXF fit in the individual bins (see \citealt{vanDerMarel94} for a similar analysis on the non-thermal nucleus of M~87). The plot shows that the underlying galaxy profile is smooth and well approximated by a shallow power-law $\Sigma(R)\propto R^{-0.22}$, as expected in a `core' elliptical \citep{lauer95,marconi01}. This confirms that the observed radial spectral variation is consistent with being entirely due to the PSF effect and not to an intrinsic change in the stellar population. The measurement of $\gamma(R)$ from the spectra allows for an extremely accurate determination of the halo of the PSF, which would be difficult to disentangle from the underlying galaxy via photometry alone.

\section{Stellar dynamical modeling}

\subsection{Geometry of the stellar spheroid}

Cen~A appears very round on the sky and shows little rotation within one \re. \citet{Wilkinson86} mapped the full velocity field of Cen~A with 71 slit positions out to 100\arcsec\ from the nucleus. They found a maximum rotation of around 40 \kms\ roughly along the direction of the major axis at ${\rm PA_{kin}}\approx35^\circ$. \citet{silge05} find a comparable maximum rotation of about 20 \kms\ within 40\arcsec. Adopting a luminosity-weighted dispersion $\sigma_e\approx138$  \kms\ \citep{silge05}, the amount of rotation can be quantified by the parameter $V_{\rm max}/\sigma_0\approx0.29$ \citep{binney78}. From the empirical relation $(\vs)_e\approx0.57\, (V_{\rm max}/\sigma_0)$ (equation~[23] of \citealt{cappellari07}), we estimate for Cen~A $(\vs)_e\approx0.17$ \citep{binney05}. The ellipticity of the galaxy within 1\re, measured from 2MASS $K$-band photometry, is very small $\varepsilon\la0.05$ \citep{silge05}. These measured values seem to place Cen~A on the region of the face-on fast rotator early-type galaxies on the \vse\ diagram (figure~11 of \citealt{cappellari07}).

Although the ellipticity of Cen~A remains low $\varepsilon\la0.1$ out to about 3\re, a well-defined photometric PA can still be determined. The major axis that we measure from the inertia ellipsoid of the surface brightness on the 2MASS $K$-band photometry \citep{Jarrett03} at large radii is ${\rm PA_{phot}}=35^\circ\pm3^\circ$ (East of North). The axis of maximum rotation that we determine from our 250mas SINFONI data is quite different at ${\rm PA_{kin}}=167^\circ\pm8^\circ$ (\refsec{sec:ppxf}). The sense of stellar rotation that we measure is in support to the determination by \citet{Wilkinson86}, and our accurate ${\rm PA_{kin}}$ is probably consistent with their determination, given their large uncertainties.

The strong kinematical misalignment of $\sim48^\circ$ is an indication that the galaxy is certainly not an axisymmetric object. This is consistent with the strong twist in the kinematical position angle of the planetary nebula system at large radii \citep{Hui95,Peng04pn}. However the galaxy is unlikely to be well described by a stationary triaxial geometry either. In fact at radii where Cen~A starts to appear more elongated, its surface brightness is dominated by stellar shells, which constitute the relics of a recent accretion event and are likely associated with the polar geometry of the strong dust lane which crosses the galaxy nucleus \citep{Malin83,Quillen93,israel98}. This shows that at those radii the galaxy has not reached an equilibrium configuration. The relatively low but ordered rotation near the nucleus, the presence of prominent stellar shells at large radii and the strong kinematical misalignment of Cen~A are reminiscent of the fast rotator S0 NGC~474 in \citet{emsellem07}. However only much more extended integral-field observations in the near-infrared, sampling up to 1\re, could conclusively reveal the dynamical status of the central regions of Cen~A.

At different radii the galaxy is expected to reach an equilibrium configuration in very different time scales, starting from the center. We estimated the circular velocity $V_c$ of Cen~A from the Multi-Gaussian Expansion \citep[MGE;][]{emsellem94}  parametrization of its $K$-band surface brightness tabulated in \citet{haring06} and the best fitting mass-to-light ratio $M/L$ we derive in this paper. It is nearly flat over two orders of magnitude in radius $V_c\sim200$ \kms\ from 4\arcsec--400\arcsec. This implies that at the edge of our SINFONI observations ($R\sim4\arcsec$) the characteristic orbital period is $T\sim2$ Myr, while it is $T\sim0.2$ Gyr at large radii where the stellar body starts to become more elongated and shells are still visible. The very short orbital period in the central regions, where the galaxy is nearly circular in projection, suggests that it has already reached an equilibrium configuration there. Given that the kinematics at a certain projected radius $R$ is mostly influenced by stellar orbits having that same characteristic radius $R$ \citep{krajnovic05}, this justifies the use of stationary dynamical models to study the central dynamics of Cen~A, even though the galaxy is still dynamically evolving at larger radii.

For this paper we will construct two types of models, to test the sensitivity of the \bh\ estimate to the assumed geometry and dynamics. We build (i) an axisymmetric orbit-based model, to reproduce in detail the kinematic observations, and (ii) a simple anisotropic spherical Jeans model to qualitatively check the \bh\ determination.

\subsection{Axisymmetric three-integral models}
\label{sec:3i-model}

Our axisymmetric three-integral dynamical model is based on \citet{schwarzschild79} numerical orbit-superposition method. This has become the current standard for all \bh\ determinations from the stellar kinematics, available in the literature from different groups \citep[e.g.][]{vanDerMarel98,gebhardt03,Valluri05}. The axisymmetric implementation we use, the orbital `dithering' and the setup we employ, are described in detail in \citet{cappellari06}. The relatively limited spatial extension of the integral-field data, and the likely lack of equilibrium at large radii, does not justify the use of more general triaxial models \citep{deLorenzi07,vanDenBosch08}.

Our dynamical model assumes constant $M/L$. This is likely a good approximation in the central regions ($R\la\re$) of Cen~A, where dark matter is expected to contribute only a small fraction of the mass \citep{Hui95,Peng04pn} as observed via dynamics or gravitational lensing in larger samples of early-type galaxies \citep{gerhard01,Rusin03,cappellari06,koopmans06,thomas07,Bolton08}. The model still allows for dark matter in the form of a constant shift in the global $M/L$. Our model assumes an MGE parametrization for the $K$-band surface brightness of Cen~A using the values tabulated in \citet{haring06}.

The models are fitted to our integral-field SINFONI 100mas (with AO correction) and 250mas kinematics in the central regions ($R\la4\arcsec$), and to the two-slits GNIRS kinematics of \citet{silge05} at larger radii ($R\la40\arcsec$). For $R<1\arcsec$ we only used the SINFONI 100mas data in the fit, not to spoil the information contained in the high-resolution AO-assisted observations with the seeing-limited 250mas ones. Similarly we do not fit the GNIRS kinematics for $R<3\farcs5$, where we have higher quality SINFONI data. As the models are bi-symmetric by construction, the SINFONI kinematics were bi-symmetrized before the fit along ${\rm PA_{kin}}=167^\circ$. For all datasets we fit, and we compute the predictions, for the Gauss-Hermite moments up to $h_3$--$h_6$, where $h_3$ and $h_4$ are the measured values, while we set $h_5=h_6=0\pm0.3$ ($1\sigma$ errors). This is done to include in the model the additional observational constraint that extreme values for the high-order moments are never observed (see two extreme examples in fig.~4 of \citealt{cappellari07}).

We constructed a set of models with slightly different geometries to test the variation in the recovered \bh: (i) An edge-on model ($i=90^\circ$); (ii)  An edge-on model with the major axis GNIRS kinematics aligned with ${\rm PA_{kin}}=167^\circ$. In this case we set $V=h_3=0$ for the GNIRS minor axis data; (iii) An edge-on model with `circularized' GNIRS kinematics. Here we assumed the $\sigma$ and $h_4$ field have circular symmetry, and a profile given by the average of the two slits. We fitted the $\sigma$ and $h_4$ GNIRS kinematics along five polar sectors, covering the whole galaxy from the major to the minor axis, and not just the observed major/minor axes. (iv) Same as model [ii], but with $i=60^\circ$, which is the average inclination for random orientations; (v) Same as model [ii], but with $i=45^\circ$. All these models are expected to produce similar results given that the differences between the major and minor axis GNIRS kinematics are almost at the level of the measurement errors, and given that Cen~A is so close to circular within the region where we have kinematics, that for large ranges of inclination the intrinsic density is essentially spherical. We find that indeed all these models produce BH mass estimates well within the errors, and have similar uncertainties. The orbital distribution of all the models is also nearly the same. For this reason we will not present the individual results. Instead for all our figures below we adopt as representative the model (ii).

\begin{figure}
  \plotone{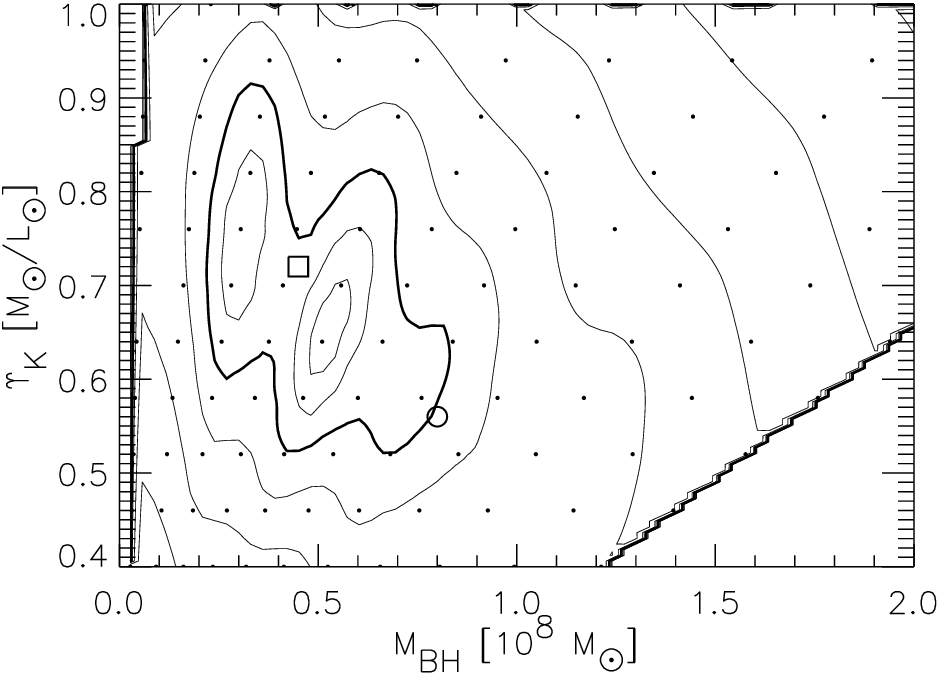}
  \caption{Determination of the best fitting parameters. The contours of the $\chi^2$, which describes the agreement between the data and the model, are plotted as a function of \bh\ and $K$-band $M/L$. The lowest levels show the $\Delta\chi^2=\chi^2-\chi^2_{\rm min}$ which correspond to confidence levels of 1, 2 and $3\sigma$ (thick contour), for one parameter. Additional contours are separated by factors of two in $\Delta\chi^2$. The open square marks the best fitting \bh\ derived from the gas kinematics by \citet{Neumayer07}. The open circle is the best fitting isotropic model from \refsec{sec:jeans}. The small dots are the locations of the models that were run, while the $\chi^2$ was interpolated at other locations.}
  \label{fig:chi2_grid}
\end{figure}

The two best-fitting parameters of the model, \bh\ and $M/L$, were determined as follows. After an initial approximate search for the best fitting parameters using isotropic Jeans models (\refsec{sec:jeans}), we computed a set of orbit libraries, each consisting of 444,528 individual orbits, at our first guess for $M/L$. For each set the \bh\ was sampled linearly in $u=\sinh(\bh/M_{\rm BH,best})$, where $M_{\rm BH,best}$ is our first guess for \bh. We then scaled the velocities of the orbits to compute models at different $M/L$ \citep{vanDerMarel98}. In the fit we use a modest amount of regularization $\Delta=10$ \citep[see][for a definition]{vanDerMarel98}. The contours of the resulting $\chi^2$, which describe the agreement between the models and the data, are shown in \reffig{fig:chi2_grid}. The best fitting BH has mass $\bh=(5.5\pm3.0)\times10^7 \msun$, where the $3\sigma$ error bars are marginalized over the $M/L$ and correspond to one degree-of-freedom ($\Delta\chi^2=9$). The corresponding best fitting $K$-band $M/L=(0.65\pm0.15)\, \msun/\lsun$.

\begin{figure}
\plotone{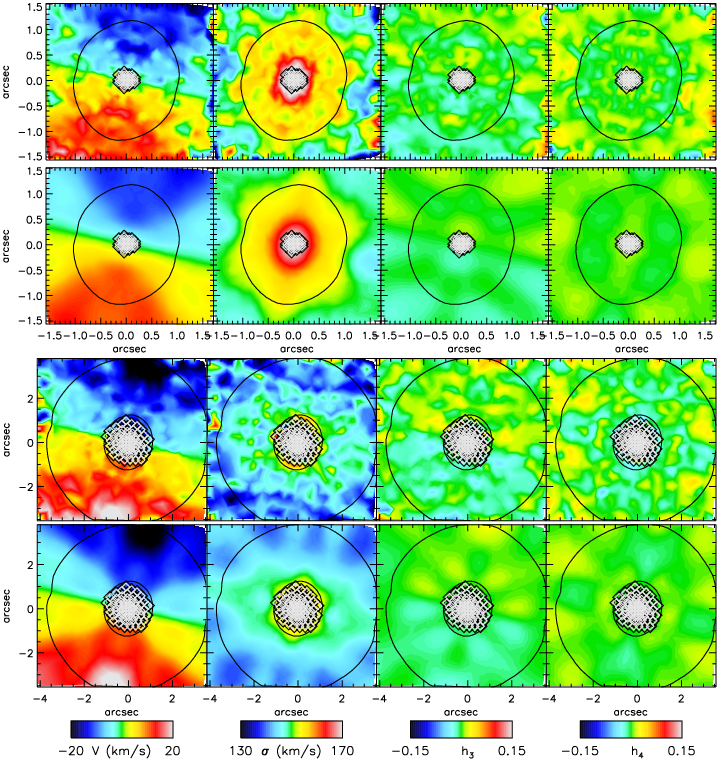}
 \caption{Data-model comparison for the best fitting three integral model. {\em Top two panels:} The top row shows the symmetrized and linearly interpolated 100mas SINFONI data of \reffig{fig:kinematics100}. The second row show the best fitting dynamical model predictions. The central bins that were excluded from the fit are shown with the white diamonds. {\em Bottom two panels:} Same as in the top two panels, for the 250mas SINFONI kinematics. For each quantity the color scale is the same in the two instrumental configurations.}
  \label{fig:best_fit_kinematics}
\end{figure}

The data-model comparison for our best fitting model are shown in \reffig{fig:best_fit_kinematics}. All the significant features of the data are reproduced in detail by the model, for the two SINFONI datasets. Unsurprisingly the GNIRS long-slit data are also reproduced in detail (not shown). Although we only have observations along two slits, we still predict the model kinematics over a full quadrant, to verify that the model does not contain un-physical features or sharp discontinuities, which are never observed in real galaxies \citep{emsellem04}.

\begin{figure}
  \plotone{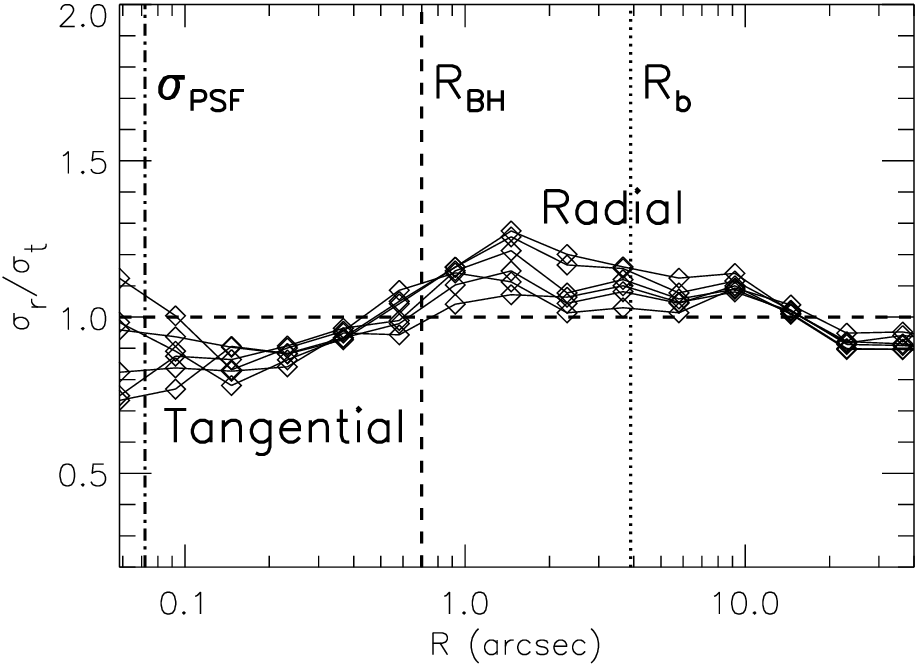}
  \caption{Anisotropy variation. The diamonds, connected with solid lines, show the anisotropy $\sigma_r/\sigma_t$ (see text for a definition) measured at different polar angles, from the equatorial plane to the symmetry axis, in the galaxy meridional plane. As the model is nearly spherical, the differences at a given radius, provide a rough indication of the model uncertainties. Also shown are the break radius (or core radius) $R_b$, the BH radius of influence $R_{\rm BH}$ and the $\sigma$ of the best fitting Gaussian model of the SINFONI 100mas PSF.}
  \label{fig:anisotropy}
\end{figure}

The orbital anisotropy of our best fitting model is shown in \reffig{fig:anisotropy}. We plot the ratio $\sigma_r/\sigma_t$ between the radial velocity dispersion $\sigma_r$ and the tangential one defined as $\sigma_t^2=(\sigma^2_\phi+\sigma^2_\theta)/2$. Here $\sigma_\phi$ includes only random motion and not ordered rotation, so that an isotropic system has $\sigma_r/\sigma_t=1$, and $(r,\theta,\phi)$ are the standard spherical coordinates. We also plot for reference the radius at which the total luminous mass in the MGE model is equal to \bh, for our best fitting $M/L$ and \bh\ values. The derived value is $R_{\rm BH}=0\farcs70$, which is very close to the usual practical definition of the BH radius of influence $R_{\rm BH}=G\bh/\sigma^2_e=0\farcs68$, for $\sigma_e=138$  \kms\ \citep[see also][]{marconi06}. Also shown is the break radius $R_b=3\farcs9$ of a Nuker-law \citep{lauer95} fit to the $K$-band HST photometry \citep{marconi00}. Both radii are very well resolved by the observations, which then allow the nuclear orbital distribution to be robustly recovered. The velocity ellipsoid appears to deviate from a sphere (isotropy) by just 15\% in the whole range where we have kinematics. This is just above the errors of 5\% in this determination \citep{cappellari07}. The main significant trend is a decrease in the $\sigma_r/\sigma_t$ ratio inside $R_{\rm BH}$, but no sharp transition is seen near $R_b$.

\subsection{Anisotropic spherical Jeans models}
\label{sec:jeans}

In the previous section we used general axisymmetric Schwarz\-schild models to determine \bh\ and to derive the anisotropy profile in the nucleus of Cen~A. In this section we use simpler anisotropic spherical Jeans models to test the reliability of these  modeling results.

In the past 25 years the dynamical models of galaxies have evolved from simple spherical models based on the Jeans equations \citep{binney82} to quite general orbit- or particle-based numerical techniques with spherical \citep{richstone88,rix97}, axisymmetric \citep[e.g.][]{vanDerMarel98} or triaxial geometry \citep{deLorenzi07,vanDenBosch08}. Contrary to the models based on the Jeans equations, which can generally be solved and tested to machine precision, the results of the more general models are sensitive to the numerical implementation details and more difficult to test thoroughly. Moreover the complexity of the general models often prevents a qualitative assessment of the results, which sometimes have to rely entirely on subtle details of the kinematics (e.g.\ fig.~2 of \citealt{gebhardt03} or fig.~7 of \citealt{nowak07}).

One way to gain confidence in the \bh\ determinations and the recovered anisotropy is by comparing them with the simpler Jeans models \citep{vanDerMarel98,cretton99ngc4342}. These models are based on a completely different set of assumptions and do not suffer from possible incompleteness in the sampling of the orbit library or from limited kinematic coverage. To test our best fitting \bh\ and $M/L$ and the corresponding recovered anisotropy profile of \reffig{fig:anisotropy} we use the Jeans Anisotropic MGE (\textsc{JAM}) package$^3$ of \citet{Cappellari08}. The method allows one to compute the predicted second moments, projected onto the sky plane, for a model with a variable anisotropy profile, via a quick and accurate single numerical quadrature. Given that our standard model of Cen~A is intrinsically close to spherical, we use the spherical formalism of equation~(50) of \citet{Cappellari08}.

We construct a spherical model having the same MGE surface brightness as our Schwarzschild model, but we set the axial ratios of all the Gaussians $q'_i=1$. We assume the same anisotropy profile of our best fitting Schwarzschild model of \refsec{sec:3i-model}, with the same \bh\ and $M/L$. The method allows for a different anisotropy $\beta=1-\sigma^2_\theta/\sigma^2_r$ to be assigned to different Gaussian components of the MGE model. In practice we assign an anisotropy $\beta=0.2$ (radial anisotropy) to the Gaussians with $0\farcs5<\sigma_i<10\arcsec$ and an anisotropy $\beta=-0.35$ (tangential anisotropy) to the remaining Gaussians. Although $\beta$ changes in a discontinuous fashion for the different MGE Gaussians, the resulting $\beta$ profile is smooth. In fact the Gaussians overlap with each other and the $\beta$ at a certain spatial position is a luminosity-weighted sum of the $\beta$ of the individual components.

\begin{figure}
  \plotone{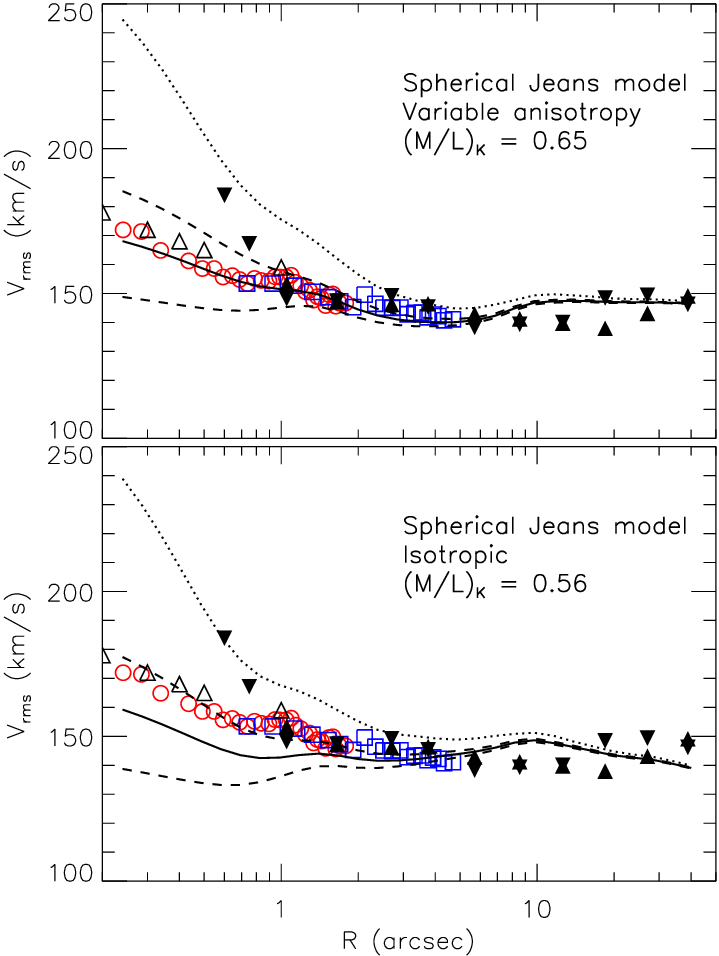}
  \caption{Spherical Jeans models. {\em Top Panel:} Our SINFONI observations of $V_{\rm rms}\equiv\sqrt{V^2+\sigma^2}$ for the 100mas (red open circles) and 250mas scale (blue open squares), biweight averaged over circular annuli, are plotted as a function of the radius from the center of Cen~A. This is the same kinematics presented in \reffig{fig:kinematics250} \ref{fig:kinematics100}, which was used in the Schwarzschild models. We also show for comparison, as black open upward triangles, the $\sigma$ measured in \reffig{fig:spectrum_variation}, assuming a purely Gaussian LOSVD.
  The corresponding GNIRS $V_{\rm rms}$ observations of \citet{silge05} along the major axis (filled downward triangle) and the minor axis (filled upward triangle), are also shown for $R\ga0\farcs5$ (see text for a discussion of the discrepancy between the GNIRS and SINFONI data inside $R\la1\arcsec$). These have been arbitrarily scaled up by 5\% to match our data at $R>1\arcsec$ (see text). The solid line shows the prediction of a \textsc{JAM} model, having the best fitting \bh\ and \ml\ derived from the Schwarzschild model of \reffig{fig:chi2_grid} and the corresponding anisotropy profile of \reffig{fig:anisotropy}. The model was convolved with the PSF of the 100mas observations. The two dashed lines have a \bh\ corresponding to the upper and lower $3\sigma$ errors in the Schwarzschild model. The dotted line has $\bh=2\times10^8$, corresponding to the best fitting value of \citet{silge05}. {\em Bottom Panel:} Same as in the top panel, but assuming isotropy ($\beta=0$) for all the Jeans models. The \ml\ has been decreased to match the data at large radii.}
  \label{fig:jeans}
\end{figure}

This anisotropic model, mimicking the best-fitting three-integral model,  is shown with the solid line in the top panel of \reffig{fig:jeans}. Although the model is not by itself a fit to the data, it provides a remarkably accurate description of the observed and radially-averaged SINFONI and GNIRS (outside $R\ga1\arcsec$) second velocity moments $V_{\rm rms}\equiv\sqrt{V^2+\sigma^2}$. We ignore the $h_4$ Gauss-Hermite moment in the $V_{\rm rms}$ estimation, as it is essentially zero within the errors. The purely Gaussian LOSVD extraction of \reffig{fig:spectrum_variation} gives consistent $V_{\rm rms}$ values within the errors. This test shows that the best-fitting discrete Schwarzschild's numerical representation of Cen~A is an excellent approximation for the continuum limit represented by the Jeans model. It provides a strong confirmation for the recovered anisotropy profile and the need for tangential anisotropy in the nucleus of Cen~A. We also show in \reffig{fig:jeans}, with the two dashed lines, the Jeans predictions for two \bh\ corresponding to the $3\sigma$ upper and lower confidence limits of the Schwarzschild model. They show that, as expected, inside $R\la1\arcsec$ the data are very sensitive to a change in \bh\ at the level of the quoted errors.

The dotted line shows the model prediction, still with the same $\beta$ profile as for the previous models, for $\bh=2\times10^8 \msun$. This is the best fitting value determined by \citet{silge05} from the GNIRS data. This model has a much steeper $V_{\rm rms}$ profile near the center, which seems to qualitatively reproduce the steeper rise of the $V_{\rm rms}$ in the GNIRS data within $R\la1$. Unfortunately we cannot compare our Jeans models with the more nuclear GNIRS kinematics ($R\la0\farcs5$). Those measurements have rather extreme $h_4\ga0.2$ values. In this case it becomes not possible to reliably translate the kinematics into a true $V_{\rm rms}$ value, as required by the Jeans equations. In fact when the Gauss-Hermite moments are large, the $V_{\rm rms}$ one can derive by formally integrating over the LOSVD is extremely sensitive to the wings of the LOSVD which are observationally not well constrained. For this reason we limit our comparison to the remaining values, for which $h_4$ is consistent with zero within the $1\sigma$ errors. Our comparison suggests that the $\sim4\times$ higher value for \bh\ derived by those authors is not due to differences in the modeling, but more likely to the differences in the measured nuclear kinematics profile with GNIRS. A more conclusive evidence would require a full Schwarzschild model of the nuclear GNIRS data, which is outside the scope of this paper.

It is clear however that our SINFONI nuclear kinematics is not consistent with the GNIRS one of \citet{silge05} when $R\la1\arcsec$. Those authors measure larger $\sigma$ values and especially strong $h_4$ values in the nucleus, while we find $h_4\sim0$ over the full field-of-view (\reffig{fig:kinematics100}). The reasons for these differences are likely the following: (i) Our SINFONI data represent a significant improvement over the GNIRS one, both in $S/N$ and in spatial resolution (compare fig.~6 of \citealt{silge05} with the right column of our \reffig{fig:spectrum_variation}). The GNIRS data forced them to extract the kinematics inside $R<1\arcsec$ by fitting only the first CO band, while we could accurately fit all three CO bands from the SINFONI data, down to the limit where they disappear due to the rise of the non-thermal nucleus; (ii) We used different approaches for the subtraction of the non-thermal continuum. As discussed in detail in \refsec{sec:nonthermal}, a non accurate subtraction of the non-thermal contribution from the spectrum, leads directly to an error in $\sigma$. We subtracted the non-thermal source with additive polynomials during the pPXF fit, so that both the kinematics and the continuum contribution are simultaneously optimized to fit the spectrum. \citet{silge05} instead removed the continuum based on a measure of CO line-strength and only subsequently fitted the kinematics on the continuum-free spectrum.

For comparison, in the bottom panel of \reffig{fig:jeans}, we show four spherical Jeans models with the same \bh\ as in the top panel, but in this case assuming isotropy ($\beta=0$). The $M/L$ was decreased to fit the data at large radii, where the BH contribution is minimal. A reasonable fit to the $V_{\rm rms}$ is obtained in this case at the upper limit $\bh=8\times10^7 \msun$ of our Schwarzschild confidence interval. This general agreement confirms previous comparisons, showing that although general three-integral models are needed for accurate \bh\ determinations, much simpler isotropic models, applied to the same data, provide a useful sanity-check for the determination \citep{vanDerMarel98,cretton99ngc4342,joseph01,verolme02}. These simple tests would be especially useful to give confidence in the \bh\ determinations in critical cases, where $R_{\rm BH}$ is barely resolved and a determination of \bh\ seems at the limit of what is possible with current technologies \citep[e.g.][]{Davies06,nowak07,Nowak08}.

\citet{Gebhardt04} compared the \bh\ determinations between two-integral and three-integral models for 10 galaxies and found that, even using seeing-limited ground-based data, the two-integral \bh\ of \citet{magorrian98} are systematically too high by just a factor $\sim2$, with respect to more recent three-integral models, based on high-resolution HST observations. It would be interesting to perform a similar type of comparison, but using the same data for both anisotropic Jeans and Schwarzschild models.

The reason for the reasonable agreement between the simpler and the more general models is likely due to the fact that many real galaxies do not `utilize' all the freedom in the orbital distribution that would be required to produce more dramatic disagreements, and do not deviate very strongly from being isotropic in their centers \citep[e.g.][]{verolme02,gebhardt03,shapiro06}. This is certainly true for the results presented in this paper.

\section{Discussion}

Our determination, using the stellar kinematics, for the BH mass in the center of Cen~A $\bh=(5.5\pm3.0)\times10^7 \msun$ agrees very well with the determination $\bh=(4.9\pm1.4)\times10^7 \msun$ we performed using the gas kinematics in \citet{Neumayer07} from the same SINFONI data. This gas-stars \bh\ comparison constitutes one of the most robust and accurate ones, due to a very well resolved BH sphere of influence ($R_{\rm BH}\approx0\farcs70$ compared to a PSF FWHM$\approx0\farcs17$) and thanks to the use of high-resolution integral-field data for both kinematical tracers. Comparable and equally successful comparisons were done by \citet{shapiro06} and \citet{Siopis08}, while a less good agreement was found in \citet{cappellari02}, likely due to the disturbed gas kinematics.

In \citet{Neumayer07} we made a detailed comparison with all the numerous previous \bh\ determinations for Cen~A \citep{marconi01,marconi06,silge05,haring06,krajnovic07}. In summary the only significant disagreement in \bh\ is with the previous stellar \bh\ determination by \citet{silge05}. In \refsec{sec:jeans} we showed that the disagreement is likely caused by a difference in the data quality and in the treatment of the contribution from the central non-thermal continuum in the kinematic extraction. It is not due to differences in the details of the modeling methods. As shown in \citet{Neumayer07}, with our new gas and stars \bh\ determination Cen~A lies within the errors on the $\bh-\sigma$ relation as given by either \citet{tremaine02} or \citet{ferrarese05rev}.

The value $(\ml)_K=0.65$ is well consistent with what one would expect from a $\sim7$ Gyr luminosity-weighted age of the stellar population of Cen~A, almost independently of the assumed near-solar metallicity. This adopts the \citet{kroupa01} {\em normalization} of the initial mass function (IMF) and uses the population models of \citet{maraston05}. These population models include a proper treatment of the TP-AGB stars which is essential for a reliable prediction of the near-infrared flux and $(\ml)_K$. The $(\ml)_K$ of the stellar population would decrease to 50\% of the observed dynamical value if Cen~A had a mean age of $\sim3$ Gyr. In that case dark-matter would be required to explain the observations. However the younger age is at the lower extreme of the estimated age of 3--8 Gyr, for the main body of Cen~A, from the analysis of its globular cluster system \citep{Peng04gc}. Adopting the normalization of the \citet{salpeter55} IMF, the mean age of Cen~A would have to be lower than $\sim3$ Gyr for the population $(\ml)_K$ not to over-predict the dynamical one. However this IMF normalization has been shown to be inconsistent with the observed dynamical \ml\ of early-type galaxies as a class (\citealt{cappellari06}; see \citealt{deJong07} for a review).

Cen~A possesses a clear core in the luminosity profile, with a break from a shallow central slope to a steeper outer one at a radius $R_b\approx3\farcs9$ \citep{marconi00}, which is extremely well resolved by our kinematics. Every galaxy spheroid seems to host a supermassive BH \citep{magorrian98,ferrarese00,gebhardt00}, and galaxies are expected to form by mergers. These cores have been interpreted as due to the scouring of the galaxy profile due to the ejection of stars in radial orbits passing close to the BH binary which forms shortly after mergers \citep{faber97,quinlan97,milosavljevic01}. The models predicts that the anisotropy should show a bias towards tangential orbits inside $R_b$.

To test this prediction, in this paper we present the recovered stellar anisotropy profile $\sigma_r/\sigma_t$ for Cen~A (\reffig{fig:anisotropy}). The profile is to first order flat around the value $\sigma_r/\sigma_t\approx1$. The maximum deviations of the velocity ellipsoid from a sphere (isotropy) are on the order of 15\%. The profile shows no sudden bias towards tangential orbits inside $R_b$, while a tangential bias appears only inside $R_{\rm BH}$. This profile seems typical of the few nuclear anisotropy profiles that have been published so far using integral-field observations or spherical models (the only cases for which the orbital distribution can be robustly recovered): M32 \citep{verolme02}, M87 \citep{cappellari05}, NGC~3379 \citep{shapiro06}, NGC~1399 \citep{houghton06,Gebhardt07}. A similar anisotropy trend was observed with high-resolution HST long-slit data by \citet{gebhardt03}, although in that case the profiles appear more noisy.

Although the predicted tangential bias is observed, its size does not seem related to that of the core. So these anisotropy determinations would not seem to support the scenario in which the cores have been produced by the core-scouring mechanism. However a big caveat is that the current simulations are not entirely realistic: they only consider binary mergers of cusp galaxies which are initially isotropic. Now that high-quality anisotropy measurements are starting to appear, it would be valuable if the numerical simulations could revisit their predictions in a more cosmological motivated setting, considering sequences of multiple mergers. For the moment we must refrain from making firm statements on what our anisotropy profile implies for BH formation.

\section{Summary}

We present a determination of the mass of the supermassive BH in the nucleus of the nearby elliptical galaxy Cen~A, using high spatial resolution integral-field observations of the stellar kinematics, combined with large-scale long-slit kinematics. Our high-resolution observations were obtained in the near infrared $K$-band with SINFONI on the VLT, using adaptive optics to correct for the blurring effect of the atmosphere. The PSF of our data has a close to diffraction-limited resolution of 0\farcs17 FWHM and very high $S/N\ga80$ per spectral pixel. We discuss the extraction of the stellar kinematics, with particular emphasis on the treatment of the non-thermal continuum coming from the nucleus of Cen~A, which contaminates the central stellar spectra.

We use a standard three-integral axisymmetric numerical orbit-superposition model to determine the mass of the supermassive BH, which turns out to be $\bh=(5.5\pm3.0)\times10^7 \msun$. This value is in good agreement with our previous accurate determination $\bh=(4.9\pm1.4)\times10^7 \msun$, obtained from the gas kinematics extracted from the same SINFONI data \citep{Neumayer07}. It is also in agreement with the recent analysis of the HST gas kinematics by \citet{marconi06}, and with other similar, but with lower spatial resolution, \bh\ determinations. This mass is consistent with the prediction of the $\bh-\sigma$ relation of either \citet{tremaine02} or \citet{ferrarese05rev}. We carefully test the \bh\ and anisotropy recovery from our three-integral stellar modeling using simple anisotropic spherical MGE Jeans models. We find very good consistency between the two different approaches.

We study the nuclear orbital distribution of Cen~A. In agreement with a few previous studies of other galaxies, we find a tangential bias in the anisotropy near the BH. However the size of the tangential region do not seem to be associated with that of the core radius $R_b$ of the surface brightness profile, but has the size of the much smaller BH sphere of influence $R_{\rm BH}$. This does not add support to the scenario in which galaxy cores are scoured by binary BHs. More realistic numerical simulations would be required for a comparison with the new accurate data and to draw conclusions from this observation.

\section*{Acknowledgements}

MC acknowledge support from a STFC Advanced Fellowship (PP/D005574/1).
NN acknowledges support from the Christiane-N\"usslein-Volhard Foundation. JR acknowledges financial support from the Academy of Finland (project 8121122). We are grateful to the referee Karl Gebhardt for a thoughtful report, which improved the presentation of our work. Based on observations collected at the European Southern Observatory, Paranal, Chile, ESO Program 075.B-0490(A).

\label{lastpage}

\end{document}